%% file: paper.tex
\documentclass[12pt,preprint]{aastex}

\newcommand{\msun}{\ensuremath{\mathrm{M}_{\sun}}}
\newcommand{\msunperyr}{\ensuremath{\mathrm{M}_{\sun}\,\mbox{yr}^{-1}}}

\newcommand{\figref}[1]{Figure \ref{#1}}
\newcommand{\eqref}[1]{equation (\ref{#1})}

\newcommand{\diskmod}[2]{$(#1,10^{-#2})$}
\newcommand{\sub}[1]{\ensuremath{_{\mbox{\scriptsize#1}}}}

\begin{document}
%\slugcomment{DRAFT --- \today}
\title{Constraints on the Formation of the Planet In HD 188753}
\shorttitle{PLANET FORMATION IN HD 188753}
\author{Hannah Jang-Condell}
\shortauthors{JANG-CONDELL}
\affil{Carnegie Institution of Washington, Department of Terrestrial Magnetism}
\affil{5241 Broad Branch Road NW, Washington, D.C. 20015}
\email{hannah@dtm.ciw.edu}

\begin{abstract}
The claimed discovery of a Jupiter-mass planet in the close triple
star system HD 188753 poses a problem for planet formation theory.  A
circumstellar disk around the planet's parent star would be truncated
close to the star, leaving little material available for planet
formation.  In this paper, we attempt to model a protoplanetary disk around
HD 188753A using a fairly simple $\alpha$-disk model, 
exploring a range of parameters constrained by observations
of T Tauri-type stars.  The disk is truncated to within 1.5 to 2.7 AU, 
depending on model parameters.  
We find that the in situ formation of the
planet around HD 188753A is implausible.  
\end{abstract}


\section{Introduction}

Recently, \citet{HD188753} has claimed the 
detection of a short-period Jupiter-mass planet 
in the close triple system HD 188753. 
The most luminious member, HD 188753A (1.06 M$_{\sun}$), 
hosts the planet, and its companion, 
HD 188753B, orbiting at a semi-major axis of 12.3 AU, 
is actually a spectroscopic binary itself, with total mass 1.63 M$_{\sun}$.  
While the planet is a typical ``hot Jupiter,'' with a minimum mass 
of 1.14 M$_J$ and orbital period of 3.35 days, 
its existence in such a close binary system is exceptional.
With an eccentricity of 0.5, 
the B component is 6 AU from the A component at closest approach, 
which would severely truncate a protoplanetary disk around A.\@
A study of test particle orbits indicates that the disk should be 
truncated at 1.3 AU \citep{pichardo,HD188753}.  However, 
analysis of Lindblad resonances in gaseous disks in eccentric 
binary systems suggest that the disk might extend somewhat further out, 
depending on the disk's Reynolds number 
\citep[][henceforth AL]{1994ArtymowiczLubow}.  
In any case, the circumstellar disk will be truncated well within 
6 AU.  Given such a small disk, 
could the 1.14 M$_J$ planet around HD 188753 have formed in situ?  

For the purposes of this paper, we adopt an $\alpha$-disk model 
that includes heating both from viscous accretion and stellar 
irradiation from the central star.  The advantage of using such a model
is that the properties of the disk are controlled by a handful of 
parameters, even though the microphysics of the processes are not 
fully modelled.  In the case of a circumstellar disk around an isolated star, 
these models provide good fits to observed SEDs \citep{dalessio3}.  
In a binary system, the companion object can substantially 
distort the disk, creating non-axisymmetries and spiral structures.  
Adequately modelling these phenomena requires high resolution 
hydrodynamic simulations, which are time-consuming and computationally 
intensive.  The objective of this paper is to establish limits 
on the feasibility of in situ planet formation in HD 188753 without 
resorting to such simulations, hence we adopt a simple $\alpha$-disk 
model, exploring a range of disk parameters to put limits on the 
possibility of planet formation.  
Shock heating from the generation of spiral structure in the disk 
is likely to inhibit rather than encourage planet formation
\citep{2000Nelson,2005Mayer_etal}, so showing that 
planet formation cannot occur in the more 
quiescent $\alpha$-disk model puts a strong constraint on 
planet formation in HD 188753.  

Other workers have studied the possibility that HD 188753 formed in a 
crowded stellar environment and that the planet's presence around the 
A component is a result of dynamical interactions.  
\citet{2005Pfahl} and 
\citet{2005PortegiesZwart_McMillan} agree that the most likely scenario 
is that the planet formed around A, which then swapped into 
a pre-existing hierarchical triple system.  
In this paper, we support this scenario by ruling out in situ 
formation of the planet.  

Although there is some debate about exactly how planets form, 
whether by core accretion \citep[e.g.][]{pollack_cores,chambers04}
or disk instability \citep[e.g.][]{boss97,boss00,boss01}, 
it is generally accepted that planets form out of 
disks of circumstellar material accreting onto the star.  
A truncated disk around HD 188753A puts severe limits 
on the amount of material that could be available for planet 
formation.  In this paper, we explore the possibilities of planet 
formation in such a disk.  
The disk models used in this paper 
are based on established $\alpha$-disk models for T Tauri stars, 
which are the prototypes for protoplanetary disks 
\citep{calvet,vertstruct,dalessio2,paper2}.  

In \S\ref{model}, we summarize the disk model we adopt for 
this paper and describe the range of parameters we will explore.  
In \S\ref{results}, we present the results of the parameter 
study, describe the calculation of truncation 
radii for the disks, and describe the calculation of the 
particulate content of the disks.  
In \S\ref{discussion}, we discuss our results in the context of 
both the core accretion and disk instability models for 
planet formation, including effects that might enhance 
the possilibity of planet formation.  
In \S\ref{concl}, we present our conclusions.

\section{Model Description}\label{model}

The calculation for the disk models analyzed in this paper 
is described in detail in \citet{paper1,paper2}, and 
summarized in the appendix.  We assume an $\alpha$-disk model, 
where the viscosity $\nu$ is given by 
$\nu=\alpha c_s H$ where $c_s$ is the sound speed, $H$ is the 
thermal scale height of the disk, and $\alpha$ is a dimensionless 
parameter \citep{shaksun,pringle}.  The temperature of the disk is 
set by stellar irradiation at the surface and viscous heating at the 
midplane.  The radial and vertical 
density and temperature structure of the disk 
is calculated iteratively to achieve self-consistency.  
We adopt stellar parameters of mass $M_* = 1\:\msun$, 
temperature $T_* = 4280$ K, and radius $R_* = 2.6\:\mathrm{R}_{\sun}$, 
corresponding to a 1 Myr old star with metallicity $Z=0.02$ 
\citep{siess_etal}.  

The two remaining free parameters for our disk models are 
the mass accretion rate onto the star, $\dot{M}$, 
and the viscosity parameter, $\alpha$.  
Accretion rates of T Tauri stars are calculated by subtracting template 
spectra from the observed spectra and assuming that the excess 
optical and near-ultraviolet continuum flux 
comes from the accretion shock caused by disk material falling onto the 
stellar surface \citep{gullbring}.  Typical accretion rates are around 
$\dot{M}\sim 10^{-9}-10^{-7}\:\msunperyr$.  Values for 
$\alpha$ are calculated by fitting $\alpha$-disk models to 
dust emission from disks at millimeter wavelengths, 
with a typical value of $\alpha\sim0.01$ \citep{hartmann}.  
The D/H ratio in the outer solar system suggests that 
the early solar nebula may have experienced accretion rates as large as 
$10^{-5}\msunperyr$ \citep{hersant_etal}.  
FU Ori objects may accrete as much as $10^{-4}\:\msunperyr$, but 
these are transient phenomena, lasting at most 100 years, so 
these high accretion rates are not expected 
to be sustainable in the long run 
\citep{2000CalvetHartmannStrom, 1996HartmannKenyon}.
For the sake of argument, we will include an accretion rate of 
$10^{-4}\:\msunperyr$ in our suite of models, 
bearing in mind that this would be an extreme system, not 
representative of planet-forming disks in general.  
Given these observational constraints,
we calculate a grid of disk models, 
with $\alpha$ set to 0.001, 0.01 or 0.1, and 
$\dot{M}$ set to $10^{-4}$, $10^{-5}$, $10^{-6}$, $10^{-7}$, $10^{-8}$ or 
$10^{-9}\:\msunperyr$.  We calculate the models out to 256 AU, but 
consider only the material interior to the truncation 
radius to be available for planet formation.  
We shall refer to a given disk model by the coordinate pair 
$(\alpha,\dot{M})$, so 
that Model \diskmod{0.01}{7} refers to the run with 
$\alpha = 0.01$ and $\dot{M}=10^{-7}\:\msunperyr$.

\section{Results}\label{results}

\subsection{Mass Profiles}
The mass profiles of our set of models 
are shown in \figref{disks}.  
The total disk mass is defined as the mass interior to the given radius.  
Solid, dotted, and dashed lines correspond to 
$\alpha=0.001$, 0.01, and 0.1, respectively.  
Accretion rates are indicated by symbol shape and color: 
orange hexagons for $10^{-4}$, 
red circles for $10^{-5}$, 
green triangles for $10^{-6}$, 
blue squares for $10^{-7}$, 
cyan stars for $10^{-8}$, and 
magenta asterisks for $10^{-9}\:\msunperyr$.
The points mark the truncation radii of each disk model, as 
will be discussed in \S\ref{trunccalc}.

The overall mass of the disk increases with increasing accretion rate and 
decreasing $\alpha$.  Models \diskmod{0.1}{8}, \diskmod{0.1}{9}, 
and \diskmod{0.01}{9} contain less than 10 M$_J$ within 100 AU, 
so we can rule out these disks as having too little mass to form 
a Jupiter-mass planet and ignore them for the remainder of this study.  

\subsection{Disk Truncation Radius}\label{trunccalc}

The disk around a protostar will be disrupted by the orbit of a close stellar 
companion.  For the purposes of this paper, we will assume that the pair of 
stars composing HD 188753B act dynamically as a single object.  
Supposing that planet formation takes place with the stars 
in their current orbital configuration, how much will a disk around 
HD 188753A be truncated?  One way to approach the problem is to 
calculate orbits of test particles to look for stable orbits 
\citep{pichardo}.  This method gives a truncation radius of 1.3 AU, 
regardless of disk model parameters.  

Another method is to analyze resonant torques and approximate the size 
of the disk to be where resonant and viscous torques balance (AL).  
%\citep{1994ArtymowiczLubow}.  
In this case, the disk size depends on 
the Reynolds number, 
$Re\equiv[(H/r)^2 \alpha]^{-1}$, where $r$ is the disk radius.
We take $H/r=c_s/(r\Omega_p)$, where $\Omega_p = \sqrt{GM_{\star}/r^3}$ is the 
orbital angular speed of the planet.  
Figures 7 and 8 of AL %\citet{1994ArtymowiczLubow} 
show truncation 
radii versus eccentricity and Reynolds number for binary 
mass ratios of $\mu=0.1$ and 0.3, where $\mu$ is the ratio of 
the stellar mass to the total binary mass.  
HD 188753A has an eccentricity of $e=0.5$ and $\mu=0.4$.  
Reading off values from Figures 7 and 8 of AL, we can determine 
the variation of truncation radii with Reynolds number for 
$e=0.5$ and $\mu=0.1$ or $0.3$.  We can then extrapolate between 
these two curves to find truncation radii versus Reynolds number 
for $\mu=0.4$.  

\begin{figure}
\plotone{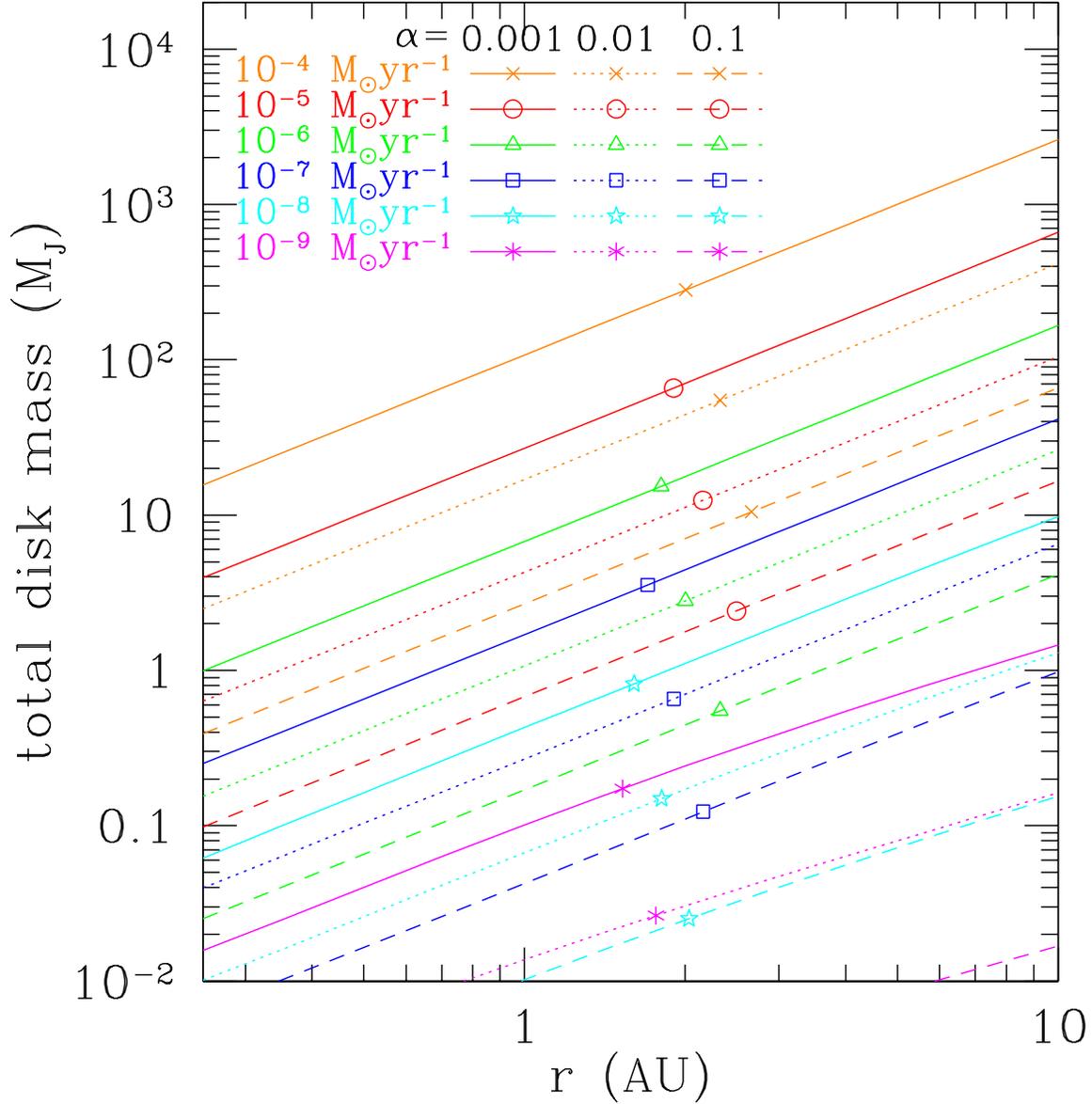} %\plotone{f1.eps}
\caption{\label{disks}
Enclosed disk mass versus radius for the suite of calculated disk models, 
in units of Jupiter masses. 
The accretion rate is indicated by color and 
symbol type:
orange hexagons for $10^{-4}$, 
red circles for $10^{-5}$,
green triangles for $10^{-6}$, 
blue squares for $10^{-7}$, 
cyan stars for $10^{-8}$, 
and magenta asterisks for $10^{-9}\:\msunperyr$.  
Models with $\alpha$ of 0.001, 0.01 and 0.01 are indicated by 
solid, dotted and dashed lines, respectively.  The locations of the points 
mark the truncation radius for each disk model.
}
\end{figure}

\begin{figure}
\plotone{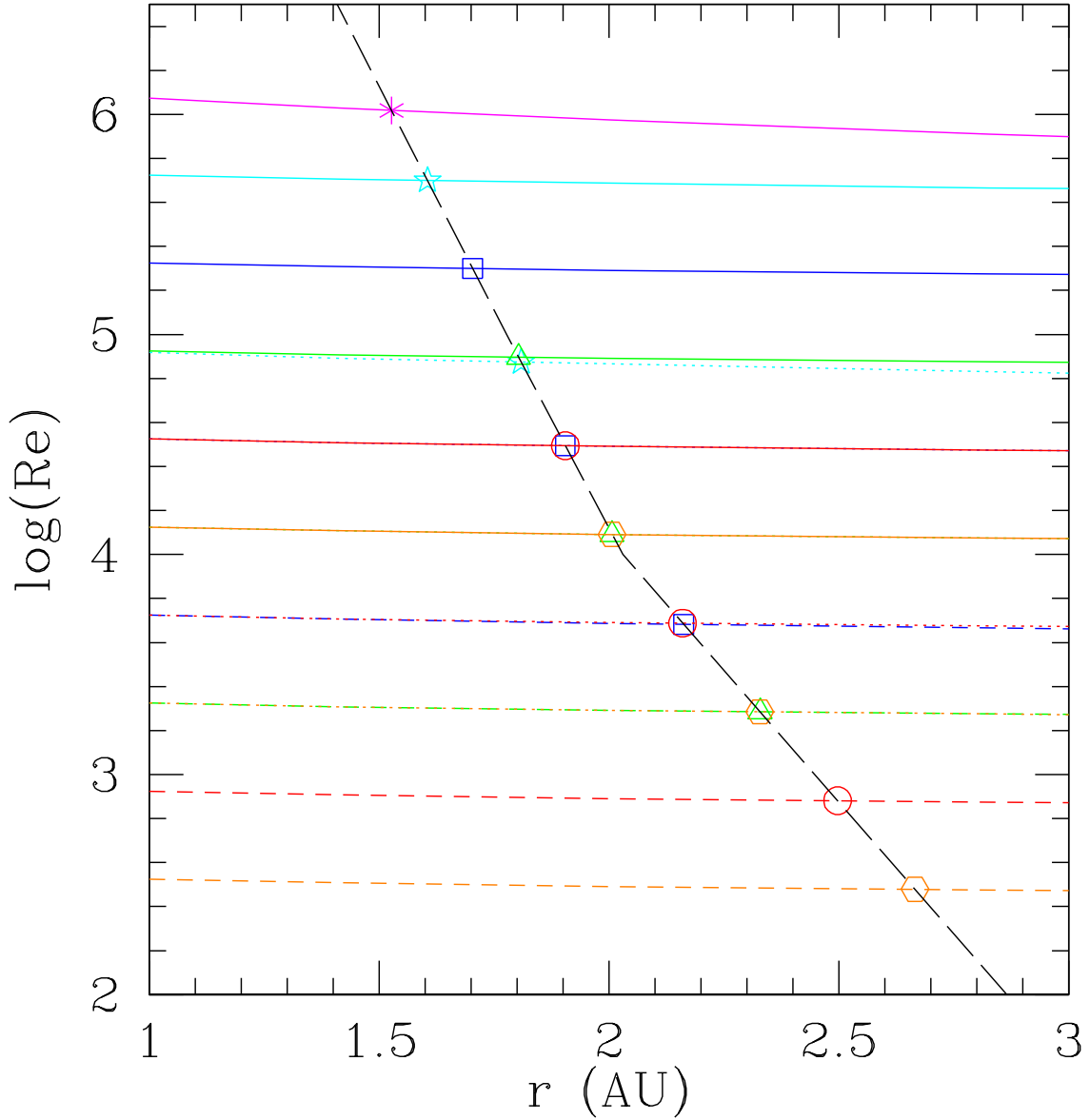} %\plotone{f2.eps}
\caption{\label{reynolds}
Reynolds number versus disk radius for our calculated disk models.  
The lines are labelled in the same way as \figref{disks}.  
The black long-dashed line illustrates the dependence of the 
truncation radius (horizontal axis) varies with Reynolds number (vertical 
axis) as calculated following AL for the parameters relevant to 
the HD 188753 system.  The intersection of this line with 
the line corresponding to each of the disk models is the truncation 
radius for that model.
}
\end{figure}

The Reynolds number in each disk model depends on the input parameters, 
but stays fairly flat with radius, as shown in \figref{reynolds}.  
The lines are labelled in the same way as in \figref{disks}.  
The dependence of truncation radius on Reynolds number as
calculated above is also plotted as a long-dashed line.  
The apparent break in slope is a result of sampling in Re value: 
AL calculated truncation radii for integral values of $\log(\mathrm{Re})$, 
so intermediate values need to be interpolated.  
From the intersection of this line with each model profile, 
we determine a unique truncation radius for each disk model.  
These radii and the enclosed disk masses ($M\sub{disk}$)
are tabulated in Table \ref{diskvals}.

These truncation radii are systematically larger than those of 
\citet{pichardo}, so we adopt the larger values to be conservative.  
We have marked the truncation radii for each model on 
\figref{disks} so we can read off the mass of disk material 
within the truncated disk.  Only disks with accretion rates of 
at least $10^{-7}\:\msunperyr$ contain more than 
1 M$_J$ within their truncation radii, ruling out at least half the models.  
Planet formation is not 100\% efficient, so we can probably 
even rule out accretion rates of less than $10^{-6}\:\msunperyr$.  

We shall assume that the disks are dynamically truncated 
and that irradiation from the stellar companion is negligible 
compared to heating from viscous accretion and the central star.  
Inclusion of this irradiation would most likely only further 
decrease the likelihood of planet formation since it would provide 
an additional heat source at the outer edge of the disk, 
inhibiting planet formation by either core accretion or disk instability.  
In the absence of additional accretion of material past the companion's 
orbit onto the disk, the disk should be viscously spreading both 
inwards and outwards.  The calculated truncated disk masses 
should be considered upper limits because of these considerations.  

\subsection{Disk Characteristics}

Table \ref{diskvals} summarizes some of the characteristics of 
out truncated disk models.  For each disk model, we list 
its truncation radius $r\sub{tr}$, the mass of the truncated disk 
$M\sub{disk}$, the disk lifetime $t\sub{disk}\equiv M\sub{disk}/\dot{M}$, 
and the minimum value for the Toomre $Q$ parameter $Q\sub{min}$.  

Disk lifetime decreases with increasing
$\alpha$ and increasing accretion rate.  However, all the disk 
lifetimes are less than $2\times10^5$ years, whereas core accretion
takes several millions of years 
\citep{pollack_cores,2003Inaba_etal,2005Hubickyj_etal}.  
The presence of a circumbinary disk may provide a reservoir that 
replenishes the circumstellar disk and extends its lifetime 
\citep{1996ArtymowiczLubow,2001WhiteGhez}.  
AL %\citet{1994ArtymowiczLubow} 
predict that accretion of 
circumbinary material should proceed faster onto the smaller-mass 
star, but observations of binary T Tauri stars appear to refute 
that \citep{2001WhiteGhez}.  

The Toomre $Q$ parameter is a measure
of whether or not a gaseous disk is locally stable to axisymmetric
perturbations.  It is defined as
\begin{equation}
Q = \frac{c_s \kappa} {\pi G \Sigma}
\end{equation}
where $\kappa$ is the epicyclic frequency, 
$G$ is the gravitational constant, 
and $\Sigma$ is the local gas surface density of disk \citep{binneytremaine}.  
In a disk with approxmate Keplerian rotation, 
$\kappa\approx\Omega_K$.
In our $\alpha$-disk models, $Q$ decreases with radius, so $Q\sub{min}$ is 
evaulated at $r\sub{tr}$.  We also find that $Q\sub{min}$ is inversely
correlated with $M\sub{disk}$ -- the most massive disks have the lowest 
values of $Q$.  The criterion for stability against 
fragmentation is $Q\gtrsim1$, so all our disks are stable.

\subsection{Solid Formation}\label{solidform}
\begin{figure}
\plotone{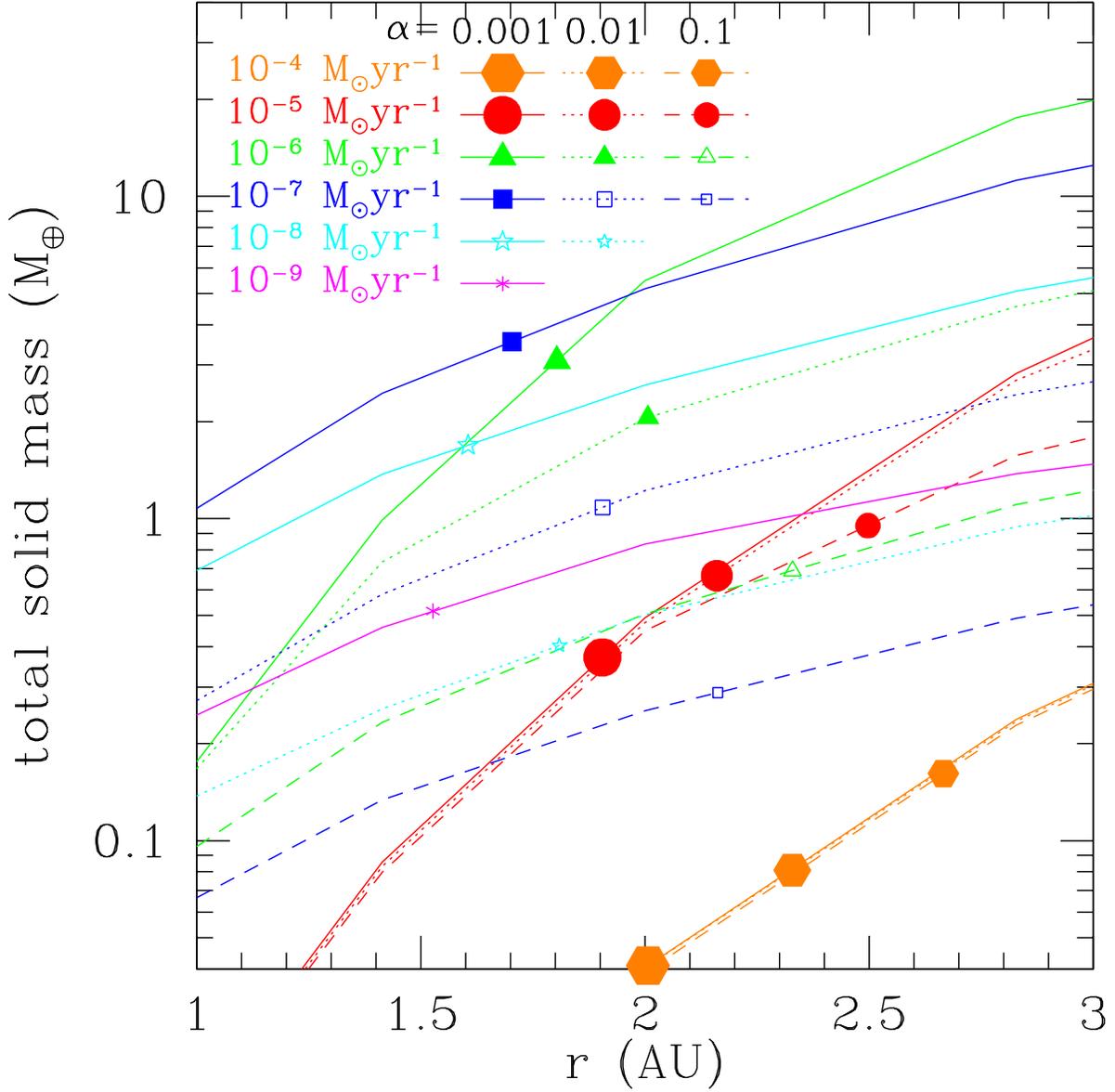} %\plotone{f3.eps}
\caption{\label{solids}
Total amount of solid material, 
in earth masses, available for 
core formation versus enclosing radius.  The lines are labelled in the 
same way as \figref{disks}.  The points mark the truncation radius 
for each disk model, with the size of the triangle indicating 
the relative overall disk mass.  Filled (open) triangles mark 
disks containing more than (less than) 1 M$_J$.
}
\end{figure}

In order to form a planet by core accretion, there must be sufficient 
mass of solid material to coagulate into a dense core which can then accrete 
gas.  We use the results of \citet{pollack_dust} for sublimation 
temperatures and mass fractions of 
olivines, orthopyroxene, iron, water, troilite, 
refractory organics and volatile organics, which compose the bulk 
of the dust in the protoplanetary disks.  We take into account 
the variation of sublimation temperatures with gas density 
and calculate the total amount of solid material 
available in the disk as a function of disk size.  
These results are plotted in \figref{solids}.  The line types correspond 
to the same models as the line types in \figref{disks}.  
The disk models were calculated at intervals of $\Delta\log(r)=\log\sqrt{2}$, 
which is evidenced as apparent breaks in slope in the disk profiles.  
The steepening of the slopes toward small $r$ is due to the sublimation 
of solids with increasing temperature.  
Truncation radii are indicated by points whose size indicate the 
relative total masses of the disks.  Filled symbols mark those disks 
with more than 1 M$_J$ total mass, open symbols (and asterisks) 
mark those below 1 M$_J$ in total mass.

In general, disks with higher accretion rates and lower $\alpha$ are hotter.  
The disks with the highest accretion rates are therefore depleted in solids, 
even though they are overall more massive.  Lower values of $\alpha$ 
favor both more massive disks as well as more solid condensation.  
Models \diskmod{0.001}{7} and \diskmod{0.001}{6} are the most favorable 
for planet formation under the core accretion scenario, but even they 
contain just a few M$_{\earth}$ of material.  
Given that at least 10 M$_{\earth}$ are required 
to form a Jupiter-mass planet, this amount
of solid material is insufficient for planet formation
\citep{pollack_cores,2003Inaba_etal,2005Hubickyj_etal}.

\section{Discussion}\label{discussion}

There are two main paradigms of planet formation -- core accretion 
and disk instability.  In core accretion, solid particles 
grow and aggregate until a body with sufficient mass to accrete 
gas forms.  In disk instability, the disk fragments into 
gas giant planets.  In this section, we address the likelihood of 
either mechanism taking place in the suite of disk models we have 
calculated.  

\subsection{Core Accretion}

As discussed in the previous section, none of our disk models 
contain enough solids within $r\sub{tr}$ to form a sufficiently 
massive core to accrete gas.  In addition, the lifetimes of the 
disks are too short for core accretion to occur.  
However, observations of young binary stars indicate that 
their disks are often replenished by a circumbinary reservoir of 
material \citep{MoninClarkePratoMcCabe}.  
Here, we will attempt to salvage core accretion by 
extending the disk's lifetime via replenishment 
and allowing additional solid material filter through the disk.
Can a 10 M$_{\earth}$ planet core still be assembled under these conditions?

Solid particles can cross the gap between a circumbinary disk 
and a circumstellar disk if they are small and coupled to the gas.  
Otherwise they do not lose angular momentum efficiently enough to 
be captured by the star.  This rules out the addition of large planetesimals 
($\gtrsim 1$ km) to the circumstellar disk that could be precursors to 
a large planet embryo -- planetesimals must be built up from 
small dust particles.  

As particle sizes increase, they experience drag forces that 
move them through the disk and may 
allow local enhancements in solid particle densities 
\citep[e.g.][]{YoudinShu,YoudinChiang}.  
When the surface density of particles reaches a critical value, 
$\Sigma_c$, the particles may be able to undergo gravitational 
instability to quickly form planetesimals.  This is not be 
confused with disk instability, which is when the gas in the disk 
fragments into gravitationally bound clumps.  Disk instability 
will be discussed later in this paper.  
We will follow the formalism of \citet{YoudinShu}, ignoring 
turbulent stresses, for simplicity.  
\citet{2003Weidenschilling} argues that turbulent stresses 
would inhibit particle accumulations, which would only further 
strengthen our result.  On the other hand, \citet{YoudinChiang} 
find that turbulent stresses should hasten accumulations rather 
than inhibit them, but since we will be 
calculating steady state particle densities, our results 
will not be affected qualitatively.  

We define $\Sigma_p$ as the surface density of solids.
When $\Sigma_p$ reaches a certain critical 
threshold, turbulent mixing can no longer prevent particles from 
settling out to the midplane.  At this point, the midplane particle 
density formally goes to infinity and gravitational instability 
of the particle layer can result in the rapid formation of large 
bodies \citep{1998Sekiya,YoudinShu}.  
This critical surface density is 
\begin{equation}
\Sigma_c = 2\sqrt{\mbox{Ri}_c} \eta r \rho_g s(\psi)
\end{equation}
where $\mbox{Ri}_c\approx 1/4$ is the Richardson number at the onset of 
Kelvin-Helmholtz instability, 
$\rho_g$ is the density of the gas at the midplane, 
$\eta$ is fractional velocity differential between the gas and 
Keplerian rotation 
\begin{equation}
\eta \equiv -\frac{\partial P/\partial r}{2\rho_g r \Omega_K^2},
\end{equation}
where $P$ is the gas pressure and
\begin{equation}
s(\psi) = (1+\psi) \ln[(1+\psi+\sqrt{1+2\psi})/\psi] - \sqrt{1+2\psi}, 
\end{equation}
where \(\psi = 4\pi G\rho_g/\Omega_K^2.\)

\begin{figure}
\plotone{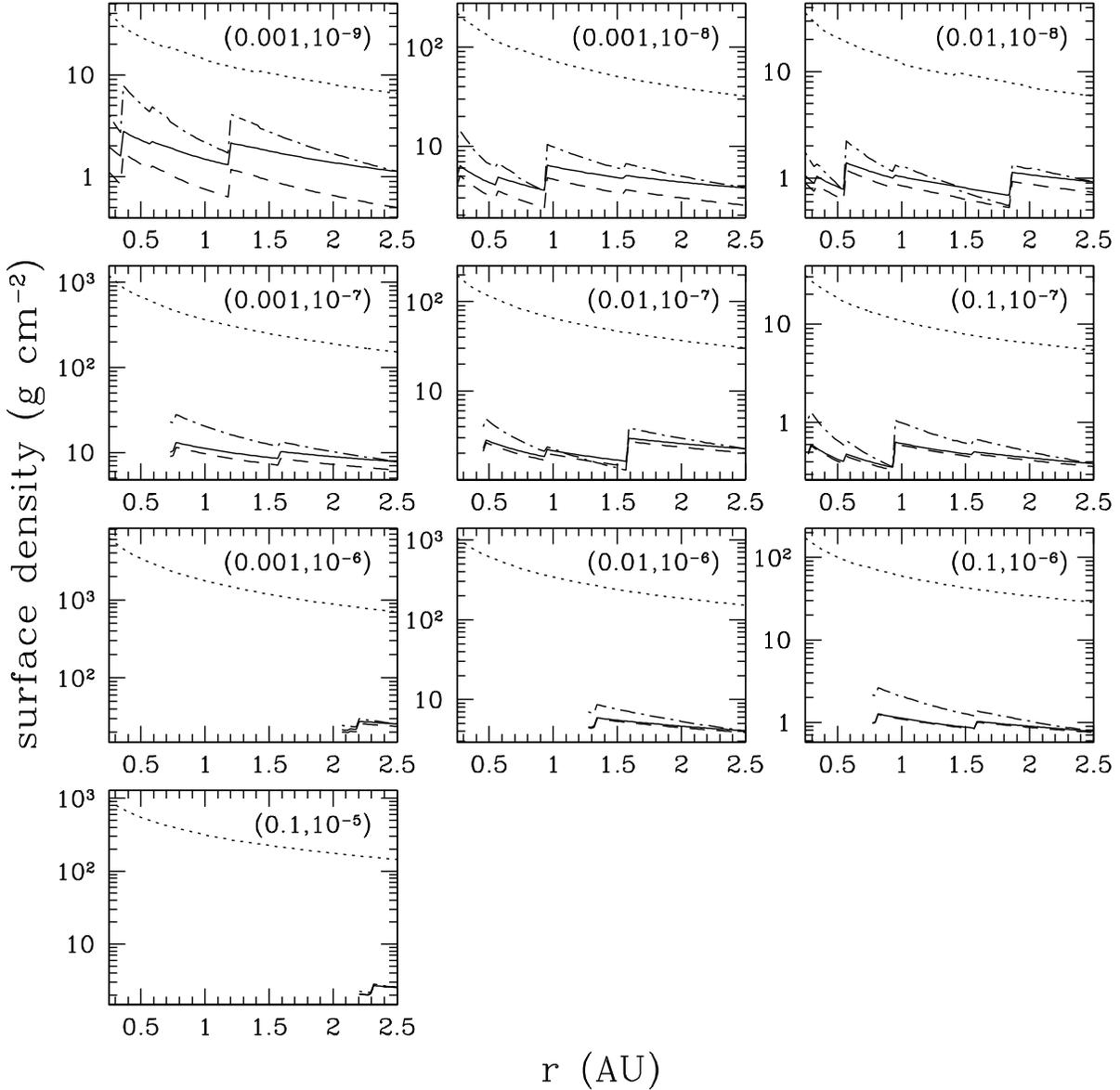} %\plotone{f4.eps}
\caption{\label{pileup}
Surface densities of solids as a function of distance from the star 
for planetesimal formation via disk instability, 
for each model disk as labelled.  
In each plot, the solid line shows the initial surface density of 
solids, $\Sigma_{p,0}$.  
The dotted line shows $\Sigma_c$, the critical surface density required for 
disk instability to occur.  
The dot-dashed line shows $\Sigma_{p,d}$, the steady-state 
surface density of solids assuming that all the particles spiral 
in with a drift rate of \(dr/dt=v\sub{Ep}\).  
The dashed line shows $\Sigma_{p,g}$, the steady-state 
surface density of solids assuming that particles drift inward 
as \(dr/dt = v\sub{gas}+v\sub{Ep}\).
The models $(0.001,10^{-5})$, $(0.01,10^{-5})$, and all those 
accreting at $10^{-4}\:\msunperyr$
are too hot for any solids to form inside of 2.5 AU.  
}
\end{figure}

In \figref{pileup}, we plot $\Sigma_c$ versus radius 
for our models as dotted lines.  For comparison, we also plot 
the unperturbed surface density of solids ($\Sigma_{p,0}$) as solid lines.  
The models are sorted column-wise by $\alpha$ as labelled 
at the top, and row-wise by accretion rate as labelled on the right.  
We have omitted \diskmod{0.001}{5} and \diskmod{0.01}{5} because the 
temperatures in those disks are everywhere too hot inside of 2.5 AU 
for any solids to remain.  In all our models, $\Sigma_{p,0}$ 
is at least an order of magnitude 
below the threshold for gravitational instability to operate.  

A particle of radius $a$ is subject to a headwind as it decouples from 
the gas and tries to orbit at the Keplerian velocity rather than 
the gas velocity, which is slower due to pressure support 
\citep{1977Weidenschilling,YoudinShu}.
The radial drift of the particle can be described by 
the Epstein drag law:
\begin{equation}
v\sub{Ep} = -\frac{t\sub{st}}{\rho_g} \frac{\partial P}{\partial r} =
2 \eta t\sub{st}
\Omega_K^2 r
\end{equation}
where the stopping time is defined as
\begin{equation}
t\sub{st} = \frac{\rho_s a}{\rho_g c_s}
\end{equation}
where $\rho_s$ is the bulk density of the particle. 

Does $\Sigma_p$ ever exceed
the critical density for gravitational instability to operate?  
To answer this question, we need to calculate how 
$\Sigma_p$ evolves according to the drift rates calculated above.  
We will assume that the gas, which composes the 
bulk of the mass in the disk, is unperturbed by the movement 
of the solid particles.  
Our assumptions will be as 
generous as possible toward the onset of gravitational instability 
in order to put firm constraints on core formation.  
We calculate the evolution of $\Sigma_p$
in each of our disk models interior to 2.5 AU, and assume 
that the amount of solids at the outer boundary is held constant
by replenishment from some external reservoir, such as a 
circumbinary disk.  

The continuity equation for $\Sigma_p$
can be written as 
\begin{equation}\label{continuity}
r \frac{\partial\Sigma_p}{\partial t} = 
\frac{\partial}{\partial r} (r \Sigma_p v\sub{r} ).
\end{equation}
where $v_r$ is the inward radial velocity of the particle.  
In steady state, the left side of \eqref{continuity} vanishes.  
In principle, the time evolution of $\Sigma_p$ may result 
in significant transient increases locally in the disk.  However, 
for the following analysis, 
we have calculated both the time-dependent and steady-state 
$\Sigma_p$ and find that the time for relaxation to steady 
state is shorter than the calculated disk lifetimes, 
and there are no significant transient 
spikes in the density profile.  

The constant of integration for the right side of \eqref{continuity} 
is essentially the mass flow rate of solids through the disk, 
which is set by the boundary condition of the 
rate of replenishment 
of solids at the outer edge of the disk,
\( dM_p/dt \equiv 2 \pi r v_r \Sigma_p \).
However, we also need to take into account the sublimation of 
solids as they fall inwards into regions of higher temperatures 
and pressures.  If we define $f(r)$ to be the equilibrium 
mass fraction of solids to total disk material 
at the midplane temperature 
and density, then 
\begin{equation}
\Sigma_p = \frac{f(r)}{2 \pi r v_r f(r\sub{o})}\frac{dM_p}{dt}.
\end{equation}
where $r\sub{o}=2.5$ AU is the outer radius of the disk.

If we assume that the steady state particle inflow rate is given 
by holding $\Sigma_p(r\sub{o})$ constant and setting $v_r = v\sub{Ep}$, 
then 
\begin{equation}\label{pilenogas}
\Sigma_{p,d} 
= \frac{r\sub{o}v\sub{Ep}(r\sub{o})}{r v\sub{Ep}} f(r)\Sigma(r\sub{o}) .
\end{equation}
where the subscript $d$ indicates that we have assumed that 
$v_r = v\sub{Ep}$.  
The quantity $v\sub{Ep}(r\sub{o})/v\sub{Ep}$ is independent of $a$, so 
the total pileup particle density integrated over $a$ depends 
only on the characteristics of the unperturbed disk, not 
on the size distribution of particles.  
In \figref{pileup}, we plot $\Sigma_{p,d}$ for each of our models 
as a dot-dashed line.  Although there is some steepening of the 
density profile toward smaller radii, it is not enough to raise 
the surface density of solids above the critical threshold for 
any disk model at any radius.  

A different way to treat the boundary condition is that 
the disk is being replenished at a rate equal to that 
accretion rate, so $dM_p/dt = f(r\sub{o}) dM/dt$.  
In this case, we need to take into account the radial velocity of the 
gas itself, which is being accreted onto the star.  
The smaller particles remain coupled to the gas, 
and leaving out the gas velocity would result in an 
unphysical accumulation of the smallest particles 
at the disk's outer edge.  For a constant 
accretion rate, the radial velocity of the gas is 
\begin{equation}
v\sub{gas} = \frac{dM/dt}{2\pi r \Sigma}
\end{equation}
and the total radial velocity of the particles is 
$v_r = v\sub{gas} + v\sub{Ep}$.  
The equation for the steady-state surface density of particles 
of size $a$ is 
\begin{equation}
\Sigma_{p,g}(a) = 
\frac{f(r)}{2\pi r (v\sub{gas}+v\sub{Ep})} \frac{dM}{dt} m(a)
\end{equation} 
where $m(a)$ is the normalized mass distribution of particles of size $a$  
and the subscript $g$ refers to the gas velocity being accounted for.  
If we assume the bulk density of the particles is constant 
and the number density
of particles with radii between $a$ and $a+da$ is 
$n(a) da \propto a^{-p} da$, then
\begin{equation}
m(a) = \frac{(4-p)a^{3-p}}{a\sub{max}^{(4-p)}-a\sub{min}^{(4-p)}}
\approx \frac{(4-p)a^{3-p}}{a\sub{max}^{(4-p)}}
\end{equation}
for $p<4$ and $a\sub{max}\gg a\sub{min}$.  
The parameters for our dust model 
are $p=3.5$ with $a\sub{max}=1$mm and 
$a\sub{min}=5\times10^{-3}\:\mu\mathrm{m}$.  
Then, 
\begin{equation}
\Sigma_{p,g}(r) = \frac{f(r)\dot{M}}{2\pi r}
[v\sub{gas} v\sub{Ep}(a\sub{max})]^{-0.5}
\left[
\arctan \sqrt{\frac{v\sub{Ep}(a\sub{max})}{v\sub{gas}}} - 
\arctan \sqrt{\frac{v\sub{Ep}(a\sub{min})}{v\sub{gas}}}
\right].
\end{equation}
In \figref{pileup}, we plot $\Sigma_{p,g}$ for each of our models 
as a dashed line.  For the higher accretion rates, $\Sigma_{p,g}$
differs little from $\Sigma_{p,0}$.  This is because the radial 
velocity of the gas is higher than the radial velocity 
of the dust, so the movement of the gas dominates the flow.  
At lower accretion rates, $\Sigma_{p,g}$ actually falls below 
$\Sigma_{p,0}$ because the dust is being swept in ahead of the gas.  
This is accompanied by a slight steepening of the profile.  
Although there is some transient accumulation of solids as 
the density profile evolves to steady state, it is small and 
never approaches $\Sigma_c$.  This analysis omits many important 
physical processes such as magnetic fields, grain growth beyond 
1 mm, and radiation pressure, but any process that would allow 
for gravitional instability to occur would have to increase 
the surface density by more than an order of magnitude.

Alternatively, kilometer-sized planetesimals can form via grain-grain 
collisions in about $10^4$ years at 1 AU in the early solar nebula, assuming 
perfect sticking in collisions \citep{1993WeidenschillingCuzzi}.  
This also assumes that they grow quickly past 1-100 cm in size, 
because particles of that size spiral into the star on the order of 
100 years \citep{1977Weidenschilling}.  
In this scenario, planetesimals form out of the dust 
that is carried in with the gas from the outer reservoir.  
Once they reach about a kilometer in size, they decouple from the 
gas.  As more material comes in from the the circumbinary reservoir, 
the population of kilometer-sized bodies grows.  They collide 
with each other and coaslesce until a planet core is built up.  
However, growth of this planet core is limited by the size of the 
feeding zone and by the amount of solids available.  
Most simulations of giant planet 
core formation in the early solar nebula are set at 4-5 AU because 
the feeding zone is sufficiently large and the solid abundance is 
enhanced by ice formation at those distances \citep[e.g.~][]{pollack_cores}.  
This is well outside the disk truncation radius for HD 188753A.  
Moreover, stirring by the eccentric binary 
companion will raise relative velocities between planetesimals, 
increasing the likelihood that collisions will result in disruption 
of bodies rather than coalescence \citep{2006ThebaultMarzariScholl}.  
A detailed simulation of this process is outside the scope of this paper, 
so we will rule planet formation in HD 188753 via core accretion 
unlikely, but not impossible.

\subsection{Disk Instability}

Since disk instability can occur within a thousand years, 
perhaps this is a more plausible mechanism for planet formation 
\citep[e.g.][]{boss01}.  
If $Q\lesssim 1$, disk instability can operate -- that is, 
the disk needs to be sufficiently cold and/or massive in order 
for gravitationally bound clumps to form.  
The minimum $Q$ is shown for our various disk models in Table \ref{diskvals}.  
For all the disk models, $Q$ is 
above the stability threshold of $Q\gtrsim1$.

\begin{figure}
\plotone{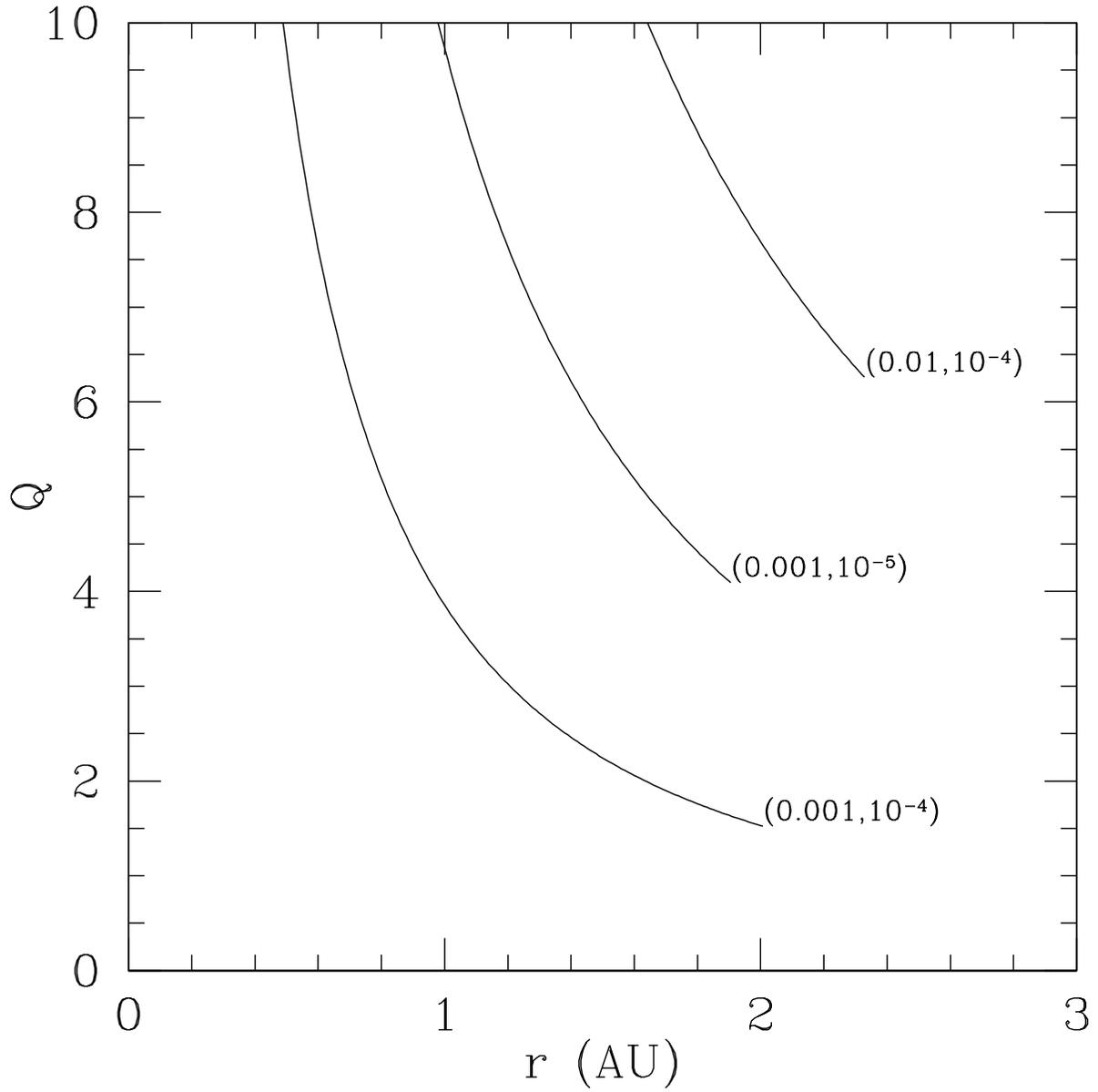} %\plotone{f5.eps}
\caption{\label{Qplot}
Toomre $Q$ parameters versus radius for those disk models with the lowest 
values of $Q$.
}
\end{figure}

\citet{boss01} shows that disk instability can act in disks 
that are only marginally stable, with $Q\approx1.5$, which 
the disk model \diskmod{0.001}{4} achieves at its outer edge.  
In \figref{Qplot}, we plot the variation of $Q$ with 
distance from the star for the disk models with the lowest values of 
$Q$.  Note that $Q$ increases rapidly as the gas moves inward as the 
disk gets accreted by the star.  
As discussed before, accretion rates of $10^{-4}\:\msunperyr$ are observed 
only as transient phenomena in young stars, lasting less than 100 years.  
Simulations show disk instability acting in several hundreds of years, 
so even disk instability may not proceed fast enough to create a 
giant planet \citep{boss01}.  
This disk model has a lifetime of only 3000 years, 
which also limits the timescale for planet formation.  
The parts of the disk that are marginally unstable at the outer edge 
accrete inward rapidly to where $Q$ is well above the stability 
threshold.  

The eccentric stellar companion to HD 188753 could excite spiral
density waves in the disk, increasing the surface density in parts of
the disk.  Since $Q$ varies inversely with $\Sigma$, local
enhancements of the surface density may make parts of the disk
unstable to gravitational collapse.  The most massive disk models also
have the smallest values of $Q$, so planets may be able to form if
density waves can create sufficient density enhancements.  
On the other hand, shock heating raises the temperature of the disks, 
which in turn raises $Q$.  

A number of simulations of triggered planet formation in binary disks 
have been carried out.  \citet{2006Boss_binary} 
shows that a binary companion can trigger disk fragmentation, whereas
\citet{2000Nelson} finds that shock heating 
inhibits disk fragmentation.  The difference between these results is 
in cooling times -- short cooling times lead to clump formation 
while long cooling times inhibit it.  
\citet{2005Mayer_etal} examine a range of cooling times for their 
simulations and conclude that shock-heating inhibits fragmentation
in general, except for the shortest cooling times.  
All these simulations were carried out in disks that were 
already only marginally stable, with $Q\sim1.4$, 
and were subject to fragmentation even in isolation, 
being cooler and larger than the ones we have modelled here.  Moreover, 
\citet{2005Mayer_etal} find that fragmentation is inhibited by the presence 
of a binary companion in more massive disks, even if they form clumps 
in isolation.  All the disks 
considered in this study are above the stability threshold
of $Q\sim1.4$.  The disks with the lowest values of $Q$ are also the most 
massive, so we conclude that the presence of the binary companion 
does not enhance disk fragmentation, 
but may even do the opposite.  Thus, we can rule out planet 
formation by disk instability.

\section{Conclusions}\label{concl}

The in situ formation of HD 188753's planet appears to be unlikely 
according to current models of planet formation.  
Assuming a simple truncated $\alpha$-disk model for the 
protoplanetary disk around HD 188753A, we have studied a 
range of disk parameters that are representative of observed 
protoplanetary disks and found that even if we can model a disk 
with sufficient total mass within the disk truncation radius, 
its timescales are too short, it contains insufficient 
solids, and it is too hot for planets to form by generally 
accepted mechanisms.  

A steady-state disk around HD 188753 does not contain enough 
solids to form a massive enough core to accrete a gaseous envelope.  
If the disk is replenished by a circumbinary reservoir, enough solids 
could be delivered to the circumstellar disk to assemble a protoplanetary 
core, in principle.  However, the temperature of the disk and the 
sitrring by the eccentric binary make it a hostile environment 
in which to assemble this core.  

Disk instability not well-favored either.  The only disk model 
that is even marginally unstable to fragmentation has an accretion
rate of $10^{-4}\:\msunperyr$, which is only seen in extreme, transient 
systems, namely FU Ori outbursts.  The instability is favored 
only in the outer reaches of the disk, which are most subject 
to perturbation and disruption by the eccentric companion.  

Admittedly, an $\alpha$-disk model is a very simple system 
and real disks are likely to be much more complex than the model
predicts.  However, it is still a useful tool for analyzing 
some of the properties of protoplanetary disks.  
A close stellar companion will create additional complexity 
to the disk model, introducing highly non-axisymmetric structures such 
as spiral waves which cannot be adequately modelled by a simple 
$\alpha$-disk.  Excitation of spiral density waves is more likely 
to disrupt planet formation than enhance it through shock heating, 
which will sublimate dust particles and elevate $Q$.  
The eccentric stellar companion is also likely to 
stir up any forming planetesimals and destroy them by smashing them 
into each other.  
To adequately model these effects, however, requires a 
high-resolution hydrodynamic simulation, which is beyond the 
scope of this paper.

\acknowledgements

Many thanks go to John Chambers, Dimitar Sasselov, Alan Boss, and
Alycia Weinberger for helpful discussions and comments in the
preparation of this paper.  
This research was supported by the NASA Astrobiology Institute under
Cooperative Agreement NNA04CC09A.

\appendix
\section{Detailed Disk Structure}\label{apdx}

The disk model has been described in detail in \citet{paper1,paper2}, 
but a summary is provided in this section.  
We adopt the formalism developed by \citet{calvet} and 
\citet{vertstruct,dalessio2}, with some simplifying assumptions.  
We use the opacities from \citet{dalessio3} using a dust model with 
parameters $a_{\mathrm{max}} = 1\,\mbox{mm}$, $T=300$ K, and $p = 3.5$, 
assuming that the dust opacities are constant throughout 
the disk.  The values for the opacities (in $\mbox{cm}^2 \mbox{g}^{-1}$)
are as follows: the Rosseland mean opacity is $\chi_R=1.91$, 
the Planck mean opacity at disk-temperature (300 K) wavelengths is $\kappa_P=0.992$,
and the Planck mean opacities at stellar-temperature (4000 K) wavelengths are 
$\chi_P^*=5.86$ and $\kappa_P^*=1.31$.  
The fraction of stellar radiation that is absorbed is represented 
by the absorption coefficient, $\alpha_{abs} = \kappa_P^*/\chi_P^*$, 
while the scattered fraction is $\sigma = 1-\alpha_{abs}$.  

We assume that the disk is locally plane parallel to decouple the 
radial and vertical dependencies of the disk properties.  
For a given radius $r$, the vertical structure is calculated as follows.  
The optical depth is given by 
\begin{equation}
\tau_d(z) = \int_z^{z_{\infty}} \chi_R \rho(z') dz'.
\end{equation}
The density and temperature are calculated assuming 
hydrostatic equilibrium,
\begin{equation}
\frac{dP}{dz} = -\rho \frac{G M_* z}{r^3}.
\end{equation}
We assume the ideal gas law, 
$P=\rho kT/\bar{m}$, where $k$ is the Boltzmann constant 
and $\bar{m}$ is the mean molecular weight of the gas, 
which we assume to be primarily molecular hydrogen.  

The temperature in the disk as a function of optical depth and angle 
of incidence of stellar radiation $\mu_s$ can be expressed as 
\begin{equation}
T(\tau_d,\mu_s) = [T_v^4(\tau_d) + T_r^4(\tau_d,\mu_s)]^{1/4}
\end{equation}
where $T_v$ and $T_r$ are temperatures due solely to viscous heating and 
stellar irradiation, respectively.

We assume that viscous flux is generated at the midplane and 
transported radiatively in a grey atmosphere so that 
\begin{equation}
T_v = \left[\frac{3F_v}{8\sigma_B}(\tau_d+2/3)\right]^{1/4}
\end{equation}
where $\sigma_B$ is the Stefan-Boltzmann constant.
The viscous flux $F_v$ at a distance $r$ for a star of 
mass $M_{\star}$ and radius $R_{\star}$ 
accreting at a rate $\dot{M}_a$ is 
\begin{equation}
F_v = \frac{3GM_{\star}\dot{M}_a}{4\pi r^3}
	\left[1-\left(\frac{R_{\star}}{r}\right)^{1/2}\right]
\end{equation}
\citep{pringle}.

Since we have assumed that the opacities are constant throughout, 
the optical depth to stellar frequencies is related to the 
optical depth to disk frequencies as $\tau_s = (\chi_P^*/\chi_R) \tau_d$.  
The equation for $T_r$ is 
\begin{equation}
\frac{\sigma_B T_r^4}{\pi} = \frac{\alpha_{abs} F_{irr} \mu_s}{4\pi}
   \left[ c_1 + c_2 e^{-\tau_s/\mu_s} + c_3 e^{-\beta\tau_s} 
     \right], 
\end{equation}
where we define $\beta = \sqrt{3\alpha_{abs}}$ to get 
\begin{eqnarray}
c_1 &=& \frac{ 6 + 9\mu_s\chi_R/\chi_P^* }{\beta^2}
- \frac{6 (1-\chi_R/\chi_P^*)\left(3-\beta^2\right)}{\beta^2
   (3 + 2 \beta) (1 + \beta \mu_s)}
\\
c_2 &=& 
\left(\frac{\chi_P^*}{\mu_s \kappa_P}-\frac{3\mu_s\chi_R}{\chi_P^*}\right)
\frac{(1-3\mu_s^2)}{(1-\beta^2 \mu_s^2)}\\
c_3 &=& \left(\frac{\beta \chi_P^*}{\kappa_P}-\frac{3\chi_R}{\chi_P^*\beta}\right)
\frac{(2+3\mu_s)(3-\beta^2)}{\beta(3+2\beta)(1-\beta^2 \mu_s^2) }.
\end{eqnarray}

The upper boundary condition is set so that 
$P(z_{\infty}) = 10^{-10}$ dyne, and we integrate the equations 
for $\tau_d$, $\rho$, and $T$ down to the midplane using some initial 
guess for $z_{\infty}$.  The other boundary condition is that we 
match the total integrated surface density 
\begin{equation}\label{surfden_int}
\Sigma = \int_{-z_{\infty}}^{z_{\infty}} \rho\, dz'
\end{equation}
with the surface density given by a steadily accreting viscous disk
\begin{equation}\label{surfden_vis}
\Sigma = \frac{\dot{M}}{3\pi\nu} 
	\left[1-\left(\frac{R_{\star}}{r}\right)^{1/2}\right] 
\end{equation}
\citep{pringle}.
We adopt a standard Shakura-Sunyaev viscosity with $\nu = \alpha c_s H$
\citep{shaksun}.
Depending on the difference between 
\eqref{surfden_int} and \eqref{surfden_vis}, we adjust our guess for 
$z_{\infty}$ until the values for $\Sigma$ converge.

The angle of incidence, $\mu_s$, depends on slope of surface $dz_s/dr$, 
where $z_s$ is the ``surface'' of the disk, where $\tau_s/\mu_s=2/3$.  
To get a self-consistent answer for $\mu_s$ we iteratively calculate 
the vertical structure of the disk at intervals of 
$\Delta\log r = \frac{1}{2}\log 2$, calculating the slope of the surface 
between intervals of $r$.  For completeness, the disk models are calculated 
out to 256 AU, even though only the inner few AU are relevant for this study.

\bibliographystyle{apj}
\bibliography{apj-jour,../planets,hd188753}
%\bibliography{paper.bbl}

\clearpage

\begin{deluxetable}{rrrrrr}
\tablecolumns{6}
\tablewidth{0pt}
\tablecaption{\label{diskvals}Calculated values for the disk models}
\tablehead{
\colhead{$\alpha$} &
\colhead{$\dot{M}$} &
\colhead{$r\sub{tr}$} &
\colhead{$M\sub{disk}$} &
\colhead{$t\sub{disk}$} &
\colhead{$Q\sub{min}$}\\
\colhead{} &
\colhead{($\msunperyr$)} &
\colhead{(AU)} &
\colhead{($\msun$)} &
\colhead{(yr)} &
\colhead{}
}
\startdata
\input{table1.tex}
\enddata
\end{deluxetable}

\clearpage

\clearpage

\clearpage

\clearpage

\clearpage

\end{document}

%% file: table1.tex
0.001 & $10^{-4}$ & 2.0 & $3.0\times10^{-1}$ & $3.0\times10^{3}$ & 1.5\\
0.001 & $10^{-5}$ & 1.9 & $7.0\times10^{-2}$ & $7.0\times10^{3}$ & 4.1\\
0.001 & $10^{-6}$ & 1.8 & $1.6\times10^{-2}$ & $1.6\times10^{4}$ & 11\\
0.001 & $10^{-7}$ & 1.7 & $3.8\times10^{-3}$ & $3.8\times10^{4}$ & 30\\
0.001 & $10^{-8}$ & 1.6 & $8.7\times10^{-4}$ & $8.7\times10^{4}$ & 83\\
0.001 & $10^{-9}$ & 1.5 & $1.9\times10^{-4}$ & $1.9\times10^{5}$ & 300\\
0.01 & $10^{-4}$ & 2.3 & $5.8\times10^{-2}$ & $5.8\times10^{2}$ & 6.3\\
0.01 & $10^{-5}$ & 2.2 & $1.3\times10^{-2}$ & $1.3\times10^{3}$ & 17\\
0.01 & $10^{-6}$ & 2.0 & $3.0\times10^{-3}$ & $3.0\times10^{3}$ & 48\\
0.01 & $10^{-7}$ & 1.9 & $7.0\times10^{-4}$ & $7.0\times10^{3}$ & 130\\
0.01 & $10^{-8}$ & 1.8 & $1.6\times10^{-4}$ & $1.6\times10^{4}$ & 380\\
0.1 & $10^{-4}$ & 2.7 & $1.1\times10^{-2}$ & $1.1\times10^{2}$ & 26\\
0.1 & $10^{-5}$ & 2.5 & $2.6\times10^{-3}$ & $2.6\times10^{2}$ & 72\\
0.1 & $10^{-6}$ & 2.3 & $5.9\times10^{-4}$ & $5.9\times10^{2}$ & 200\\
0.1 & $10^{-7}$ & 2.2 & $1.3\times10^{-4}$ & $1.3\times10^{3}$ & 560\\